\documentclass[a4paper,aps,prd,10pt,preprintnumbers,showpacs,twocolumn,groupedaddress,superscriptaddress,nofootinbib,amsmath,amssymb]{revtex4-1}
\usepackage[utf8]{inputenc}
\usepackage[T1]{fontenc}
\usepackage{cmap}
\usepackage[dvips]{graphics}

\def\imo{i}
\def\re#1{Re(#1)}
\def\im#1{Im(#1)}
\def\n{\tilde{n}}
\def\Order#1{{\cal O}\left(#1\right)}

\begin{document}
\title{Bernstein spectral method for quasinormal modes of a generic black hole spacetime and application to instability of dilaton-de Sitter solution}
\author{R. A. Konoplya}\email{roman.konoplya@gmail.com}
\affiliation{Research Centre for Theoretical Physics and Astrophysics, Institute of Physics, Silesian University in Opava, Bezručovo náměstí 13, CZ-74601 Opava, Czech Republic}
\author{A. Zhidenko} \email{olexandr.zhydenko@ufabc.edu.br}
\affiliation{Centro de Matemática, Computação e Cognição (CMCC), Universidade Federal do ABC (UFABC),\\ Rua Abolição, CEP: 09210-180, Santo André, SP, Brazil}

\begin{abstract}
We present the improved {\it Mathematica} code which computes quasinormal frequencies with the help of the Bernstein spectral method for a general class of black holes, allowing for asymptotically flat, de Sitter or anti-de Sitter asymptotic. The method is especially efficient when searching for purely imaginary and unstable modes and here it is used for detecting the instability region of a charged scalar field in the background of the charged asymptotically de Sitter dilatonic black hole. We show that the instability has superradiant nature and the dilaton field essentially influences the region of instability.
\end{abstract}
\maketitle

\section{Introduction}

Proper oscillation frequencies of black holes, {\it quasinormal modes} \cite{Kokkotas:1999bd,Berti:2009kk,Konoplya:2011qq} usually cannot be found analytically and a number of numerical or semianalytical methods are used. Simple and quick and, in a sense, automatic methods, such as the WKB one~\cite{Mashhoon:1982im,Schutz:1985zz,Iyer:1986np,Konoplya:2003ii,Konoplya:2003ii,Matyjasek:2017psv,Konoplya:2019hlu} unfortunately do not allow one to find quasinormal modes in all required cases, for example, when the real oscillation frequency is much smaller than the damping rate or zero~\cite{Konoplya:2019hlu}. The latter case includes important situations: instability which is represented by exponentially growing modes without oscillations when one is constrained by spherically symmetric solutions~\cite{Konoplya:2008yy}. Unstable modes can, in principle be detected via the Frobenius method~\cite{Leaver:1985ax}, yet it is based on trial and error procedure when searching roots of algebraic equations and there is a risk that some of the roots could be simply missed.

The method which is free of such deficiency is the time-domain integration of the wave equation at a fixed spacial coordinate~\cite{Gundlach:1993tp}. This method includes contribution of all overtones and the instability is simply shown as an exponential growth of the absolute values of the wave function at late times. However, higher overtones cannot be easily extracted from the time-domain profile within this approach and one is usually constrained by detecting two-three first overtones with reasonable accuracy. Moreover, even detecting instability may be challenging, because the instability may start at very late times after a long period of damped oscillations, as it occurs, for example, for black holes with higher curvature corrections~\cite{Konoplya:2008ix}. The latter would require time-consuming numerical integration of the wave equations until very late times.

The method which is efficient for finding purely imaginary, that is, nonoscillatory, quasinormal modes is a spectral method based on the Bernstein polynomial nonorthogonal basis functions~\cite{Fortuna:2020obg}. As was shown in~\cite{Fortuna:2020obg} while reproducing only several first overtones during a short computing time, it perfectly detects the algebraically special mode of the Schwarzschild black hole with high accuracy, which occurs at $n=8$, where $n$ is the overtone number. Another spectral method based on the Chebyshev polynomials was used to obtain a family of the purely imaginary modes of the Schwarzschild-de Sitter black holes~\cite{Jansen:2017oag}. The spectral method was also successfully employed to detect the instabilities of the rotating black holes in AdS~\cite{Monteiro:2009ke,Dias:2010,Dias:2013} and higher-dimensional rotating black holes~\cite{Dias:2009} and black strings~\cite{Dias:2022mde} (see also~\cite{Dias:2015nua} for review). This makes us expect that the Bernstein spectral method will be also useful for finding the instability region.

Here, using the initial code of \cite{Fortuna:2020obg} as a basis, we develop it to be even more user-friendly and solve two main problems related to it:
\begin{itemize}
\item The Bernstein spectral method produces a great number of roots of the matrix equation, not all of which are true quasinormal frequencies, so that only varying the size of the matrix, one can see which roots survive and distinguish them as quasinormal modes. We incorporated this procedure of the comparison of matrices into the code.
\item The code now can be automatically applied (without further adjustment) to asymptotically flat, de Sitter or anti-de Sitter spacetimes.
\end{itemize}

If the wave equation depends on a number of parameters, the problem of finding the instability region via the time-domain integration is extremely time consuming, because the integration must be performed plenty of times to cover all the possible values of the parameters. Thus, a problem of this kind could be an excellent playground for the Bernstein spectral method.

Such an example could provide the instability of a charged scalar field in the background of the charged asymptotically de Sitter black hole~\cite{Zhu:2014sya,Konoplya:2014lha}.
The region of instability in this case depends on the charges of the field and black hole, the black hole mass and the cosmological constant. While the instability of the charged scalar field in the Reissner-Nordström-de Sitter background was studied in~\cite{Zhu:2014sya,Konoplya:2014lha,Dias:2018ufh}, no such analysis was done for the dilatonic analog of this configuration.
Indeed, quasinormal modes of a neutral scalar field around the dilatonic black hole at nonzero cosmological constant was considered for the first time in~\cite{Fernando:2016ftj} with the help of the WKB approximation, while the charged scalar filed of the asymptotically flat case was considered in~\cite{Konoplya:2002ky}. Thus, to the best of our knowledge, no quasinormal modes of the charged scalar field in the background of the asymptotically de Sitter dilatonic black hole were considered so far.

Notice, that the dilaton-de Sitter black hole is interesting on its own, because of the holographic duality relating quantum gravity on
the de Sitter space and the conformal field theory~\cite{Witten:2001kn}. Then, the quasinormal modes of black holes are important not only from the point of view of observation of gravitational waves, but also as poles of the corresponding Green functions in the dual field theory, allowing to describe the relaxation of the corresponding quantum fields at finite temperature~\cite{Son:2007vk}. In the low energy limit of string theory, the Einstein action is modified, among other terms, by a scalar, dilaton, field.

Thus, our paper is two-fold: on the one hand it represents the improved Bernstein polynomial procedure for finding quasinormal modes of a broad class of black holes and, on the other hand, the code is used here for investigating the instability of the asymptotically de Sitter dilatonic black holes.

The paper is organized as follows. In section~\ref{sec:equations} we provide the basic information about the dilatonic black hole in the presence of cosmological constant. Section~\ref{sec:superradiance} is devoted to the discussion of the nature of instability -- superradiation.
Section~\ref{sec:timedomain} briefly reviews the time-domain integration which we also used as a complementary method for checking the obtained results. The Bernstein spectral method is discussed in section~\ref{sec:Bernstein}. In section~\ref{sec:results} we discuss the obtained numerical data on the instability region. Finally, in the conclusion we summarize the obtained results and mention some open problems.

\section{The basic equations}\label{sec:equations}

The well-known four-dimensional and asymptotically flat charged black-hole solutions with the dilaton field were obtained by Gibbons and Maeda~\cite{Gibbons:1987ps} and, independently, by Garfinkle~et~al.~\cite{Garfinkle:1990qj}. However, as was shown in~\cite{Poletti:1994ff}, no such extension is possible in the presence of cosmological constant with exponential coupling. Instead, the dilation potential can be represented is the sum of the cosmological constant (with some dimensionless coefficient) and two Liouville type terms~\cite{Gao:2004tu}. Here we will study this form of the dilaton potential.

The action for the dilaton gravity can be written as,
\begin{equation}
S = \int d^4 x \sqrt{-g } \left[ R - 2 \partial_{\mu} \varphi \partial^{\mu} \varphi - V(\varphi) - e^{- 2 \varphi} F_{\mu \nu} F^{ \mu \nu} \right],
\end{equation}
where $R$ is the scalar curvature, $F_{\mu \nu}$ is the Maxwell's field strength and $\varphi$ is the dilation field. The potential for the dilation field is given by $V(\varphi)$, which has the following form
\begin{equation}\label{dilaton-potential}
V(\varphi) = \frac{ 4 \Lambda}{3} + \frac{ \Lambda}{3} \left( e^{ 2 (\varphi - \varphi_0)} + e^{ - 2 (\varphi - \varphi_0)} \right).
\end{equation}
The metric of the dilaton-de Sitter black hole is~\cite{Gao:2004tu}
\begin{equation} \label{metric}
ds^2 = - f(r) dt^2 + \frac{ dr^2}{ f(r)} + R(r)^2 ( d \theta^2 + \sin^2 \theta d \phi^2),
\end{equation}
where
\begin{equation}\label{ddS}
f(r) = 1 - \frac{ 2 M} { r} - \frac{ \Lambda r}{ 3} ( r - 2 D), \quad R(r)^2 = r ( r - 2 D).
\end{equation}
Here, $\Lambda$ is the cosmological constant, $M$ is the mass and, $D$ is the dilation charge.
Notice that when $\Lambda=0$, the black hole solution in Eq.~(\ref{metric}) becomes the Gibbons-Maeda-Garfinkle-Horowitz-Strominger (GMGHS) black hole~\cite{Gibbons:1987ps,Garfinkle:1990qj}.
When $D=0$, the space-time reduces the Schwarzschild-de Sitter solution. Some basic properties of this black hole solution were studied in~\cite{Fernando:2016ftj,Benakli:2021fvv,Dehyadegari:2020tau}.

The dilation field $\varphi$, dilation charge $D$, and electric field $F_{01}$, for the above solution are given by the following relations,
\begin{eqnarray}
&&e^{2 \varphi} = e^{ 2 \varphi_0} \left( 1 - \frac{ 2 D}{r} \right),\\\nonumber
&&D = \frac{ Q^2 e^{ 2 \varphi_0} }{ 2 M}, \quad F_{01} = \frac{ Q e^{ 2 \varphi_0}}{r^2},
\end{eqnarray}
where $\varphi_0$ is the dilation field at $r\to\infty$ and $Q$ is the electric charge of the black hole.

The two largest roots of the equation $f(r)=0$ are the event $r_{+}$ and cosmological $r_{c}$ horizons. For large mass $M$ there are no horizons and the spacetime is a naked singularity. Since the metric has a singularity at $r = 2D$, the space-time becomes a black hole only if the radius of the event horizon $r_{+} > 2D$.

We shall further compare the dilaton-de Sitter black hole with the Reissner-Nordström-de Sitter black hole, $\varphi=0$,
\begin{equation}\label{RNdS}
f(r) = 1 - \frac{2 M} {r}+ \frac{Q^2} {r^2} - \frac{ \Lambda r^2}{3}, \quad R(r) = r.
\end{equation}

We designate the Cauchy, event, and cosmological horizons as $r_-$, $r_+$, and $r_c$ respectively ($r_- < r_+ < r_c$).
Then, the metric function $f(r)$ can be written as,
\begin{eqnarray}
f(r)=\frac{\Lambda}{3r^2}(r-r_+)(r-r_-)(r_c-r)(r+r_n),
\end{eqnarray}
where $r_n=r_-+r_++r_c$ for the Reissner-Nordström-de Sitter black hole (\ref{RNdS}) and $r_n=0$ for the dilaton-de Sitter black hole (\ref{ddS}).

A charged, massive scalar field $\phi$ in curved space-time obeys the Klein-Gordon equation
\begin{eqnarray}
[(\nabla^\nu-ieA^\nu)(\nabla_\nu-ieA_\nu)-\mu^2]\phi=0\;,
\end{eqnarray}
where $e$ and $\mu$ are, respectively, the charge and mass of
the field and $$A_\mu=-\frac{\delta^0_\mu Q e^{2 \varphi_0}}{r}$$ is the electromagnetic 4-potential of the black hole.

After the standard separation of angular variables with the help of spherical harmonics and introduction a new wave function $\Psi$, the above equation of motion can be reduced to the following form:
\begin{eqnarray}
-\frac{\partial^2\Psi}{\partial t^2}+\frac{\partial^2\Psi}{\partial r_\star^2}-2i\Phi\frac{\partial\Psi}{\partial t}+(\Phi^2-V)\Psi=0\;,\label{radial_eq}
\end{eqnarray}
where
$$dr_\star=\dfrac{dr}{f(r)}$$
is the tortoise coordinate, the dilaton field is
\begin{eqnarray}
\Phi(r)=\frac{eQ e^{2 \varphi_0}}{r},
\end{eqnarray}
and the effective potential has the form
\begin{equation}
V(r) = f(r) \left(\frac{ \ell ( \ell + 1) } { R^2(r)} + \frac{ f'(r) R'(r)+f(r) R''(r)}{ R(r)} + \mu^2 \right).
\label{potential}
\end{equation}

\begin{figure}
\resizebox{\linewidth}{!}{\includegraphics*{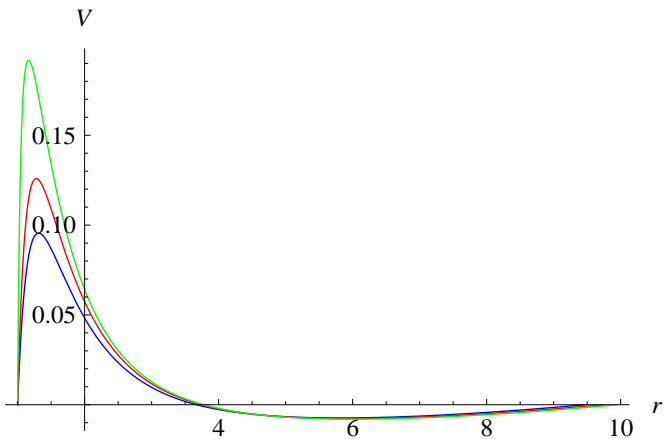}}
\caption{The effective potential $V(r)$ for $\ell=0$ case, $\Lambda = 0.03$: $D =0$ (blue, lower), $D=0.25$ (red) and $D=0.45$ (green, upper).}\label{fig:potential}
\end{figure}

The above effective potential $V(r)$ has the form of the positive definite potential barrier with a single maximum for $\ell \geq 1 $, but it has a negative gap for $\ell=0$ case (see Fig.~\ref{fig:potential}).

\section{Superradiance and instability}\label{sec:superradiance}

Here we will we consider the scattering problem for a charged scalar field in the background of the dilatonic de Sitter black hole.
Owing to the $\partial_t$ Killing vector for stationary space-times, the following ansatz $\Psi(r,t)=\exp[-\imo\omega t]\Psi(r)$ transforms (\ref{radial_eq}) to the Schrödinger-like form
\begin{eqnarray}
{\frac{d^2\Psi}{dr_\star^2}}+(\omega-\Phi(r))^2\Psi-V(r)\Psi=0\;.\label{radial}
\end{eqnarray}
Due to the following symmetry of (\ref{radial}):
$$\re{\omega}\to-\re{\omega}, \quad \Phi(r)\to-\Phi(r),$$
it is sufficient to consider only the range $eQ>0$

We will use the usual scattering boundary conditions, implying that the wave coming from the cosmological horizon will be partially reflected by the potential barrier and partially come back to the cosmological horizon, while at the event horizon of a classical black hole purely incoming wave is always required:
\begin{eqnarray}
\Psi\sim\left\{
  \begin{array}{ll}
    Be^{-\imo(\omega-\Phi(r_+))r_\star}, & r\to r_+, \\
    e^{-\imo(\omega-\Phi(r_c))r_\star}+A e^{+i(\omega-\Phi(r_c))r_\star}, & r\to r_c.
  \end{array}
\right.
\end{eqnarray}
Since the Wronskian of the complex conjugated solutions is constant, one can derive the following relation
\begin{eqnarray}
|A|^2=1-\frac{\omega-\Phi(r_+)}{\omega-\Phi(r_c)}|B|^2\;.
\end{eqnarray}
Amplification of the incident wave, that is, a \emph{superradiance}, occurs when
\begin{eqnarray}
\Phi(r_c)=\frac{eQ e^{2 \varphi_0}}{r_c}<\omega<\frac{eQ e^{2 \varphi_0}}{r_+}=\Phi(r_+).\label{regime}
\end{eqnarray}
When the cosmological constant $\Lambda$ vanishes, the superradiant condition Eq.~(\ref{regime}) is reduced to the one for asymptotically flat space-times formulated by Bekenstein~\cite{Bekenstein:1973mi,Hod:2012-2013}. Notice, that $\omega$ is the real oscillation frequency in~(\ref{regime}).

In order to understand how the condition of superradiance is related to the instability we will consider, generally, complex frequencies and impose the quasinormal boundary conditions,
\begin{eqnarray}\label{QNBC}
\Psi\sim\left\{
  \begin{array}{rcl}
    e^{-\imo(\omega-\Phi(r_+))r_\star}, \quad & r\to r_+ & (r_\star \to -\infty), \\
    e^{+\imo(\omega-\Phi(r_c))r_\star}, \quad & r\to r_c & (r_\star \to +\infty).
  \end{array}
\right.
\end{eqnarray}

Following \cite{Konoplya:2014lha} we will show that the real part of $\omega$ satisfying the superradiance inequality (\ref{regime}) is the necessary, but not sufficient, condition for the instability. For this purpose we multiply (\ref{radial}) by the complex conjugated $\Psi^\star$ and integrate the first term by parts,
\begin{eqnarray}
\Psi^\star(r_\star)\Psi^\prime(r_\star)\Biggr|_{-\infty}^\infty &+& \intop_{-\infty}^\infty (\omega-\Phi(r_\star))^2|\Psi(r_\star)|^2dr_\star \\\nonumber&=&\intop_{-\infty}^\infty\left(V(r_\star)|\Psi(r_\star)|^2 + |\Psi^\prime(r_\star)|^2\right)dr_\star.
\end{eqnarray}
The right-hand side is real, since the effective potential $V(r)$ is real. Because $\Phi(r)$ is a monotonically decreasing function, taking imaginary part of both sides, we find that if either $\re{\omega}\geq\Phi(r_+)\geq\Phi(r)$ or $\re{\omega}\leq\Phi(r_c)\leq\Phi(r)$ is satisfied, then we have $\im{\omega}<0$. This proves that the instability can take place only provided
\begin{eqnarray}\label{superradiant}
\Phi(r_c)=\frac{eQ e^{2 \varphi_0}}{r_c}<\re{\omega}<\frac{eQ e^{2 \varphi_0}}{r_+}=\Phi(r_+).
\end{eqnarray}

For $eQ e^{2 \varphi_0}$ larger than the threshold of instability the real part of the dominant mode also satisfies Eq.~(\ref{superradiant}), being damped. This is qualitatively different from the higher-dimensional asymptotically anti-de Sitter black holes, for which the necessary condition is also the sufficient one, when the black hole is charged or rotating~\cite{Kodama:2009rq,Uchikata:2011zz,Li:2012rx,Wang:2014eha}.

One should notice that the asymptotic value of the dilaton field $\phi_0$ only changes the units of the dilaton charge $D$ and electric charge of the test field $e$. Therefore, without loss of generality we shall further consider $\phi_0=0$.


When the event horizon radius is close to the de Sitter radius, that is the black hole occupies almost the whole de Sitter space, the spectral problem can be treated analytically~\cite{Cardoso:2003sw,Molina:2003ff,Churilova:2021nnc}.
In this extreme limit we expand the value of the surface gravity $\kappa$ in terms of a small difference between the radii of the event and cosmological horizons,
\begin{eqnarray}
\kappa&\equiv& \frac{f'(r_+)}{2}\\\nonumber
&=&(r_c-r_+)\frac{\Lambda(r_+ - r_-) (r_+ + r_n)}{6r_+^2}+\Order{r_c-r_+}^3.
\end{eqnarray}
Then we obtain
$$
f(r)=(r-r_+)(r_c-r)\frac{\Lambda(r_+ - r_-) (r_+ + r_n)}{3 r_+^2}+\Order{\kappa}^3,
$$
and the tortoise coordinate takes the following simple form:
\begin{equation}\label{tortoise}
r_\star\equiv\int\frac{dr}{f(r)}=\frac{1}{2\kappa}\left(\ln\left(\frac{r-r_+}{r_c-r}\right)+\Order{\kappa}\right).
\end{equation}
Therefore, one can find a closed form for $r$ in terms of $r_\star$,
\begin{equation}\label{radialexpr}
r=\frac{r_++r_c \exp(2\kappa r_\star)}{1 + \exp(2\kappa r_\star)}+\Order{\kappa}.
\end{equation}
Substituting Eq.~(\ref{radialexpr}) into the above expression for $f(r)$ we have
\begin{eqnarray}
f(r_\star)&=&\frac{(r_c-r_+)^2}{4\cosh^2(\kappa r_\star)}\frac{\Lambda(r_+ - r_-) (r_+ + r_n)}{3 r_+^2}+\Order{\kappa}^3
\nonumber\\
&=&\frac{(r_c-r_+)\kappa}{2\cosh^2(\kappa r_\star)}+\Order{\kappa}^3.
\end{eqnarray}

Using Eq.~(\ref{radialexpr}) in (\ref{potential}) we see that the potential approaches the Pöschl-Teller potential in the near-extreme limit,
\begin{equation}
V(r_\star)=\frac{A\kappa^2}{\cosh^2(\kappa r_\star)}+\Order{\kappa}^3,
\end{equation}
where the nonnegative constant $A$ is defined as follows,
\begin{equation}\label{Adef}
A=B\left(\mu^2
+\frac{\ell(\ell+1)}{R(r_+)^2}\right),
\end{equation}
and
\begin{equation}\label{Bdef}
B\equiv\lim_{r_c\to r_+}\frac{r_c-r_+}{2\kappa}=\frac{3r_+^2}{\Lambda(r_+ - r_-) (r_+ + r_n)}>0.
\end{equation}

We notice that $V(r_\star)=\Order{\kappa}^2$. Therefore, expanding the frequency in the series with respect to $\kappa$, we find that
\begin{equation}
\omega=\frac{eQ}{r_+}+\omega_{\kappa}\kappa+\Order{\kappa}^2,
\end{equation}
where $\omega_{\kappa}$ is the eigenvalue of the following equation:
\begin{equation}
\frac{d^2\Psi}{dx^2}+\left(\omega_\kappa+\frac{2BeQ}{r_e^2}\frac{e^{2x}}{1+e^{2x}}\right)^2\Psi-\frac{A}{\cosh^2(x)}\Psi=0,
\end{equation}
which is written in terms of the new coordinate
$$x\equiv \kappa r_\star.$$

Using the analytic formula for the Pöschl-Teller potential eigenvalues, we find, for small $eQ$,
\begin{equation}\label{kappaw}
\omega_\kappa=\pm\sqrt{A-\frac{1}{4}}-\imo\left(n+\frac{1}{2}\right)+\Order{eQ},
\end{equation}
implying that the scalar field in the near-extreme limit is stable for sufficiently small $eQ$, unless $A=0$, that is,
$$\im{\omega}=\kappa\im{\omega_\kappa}<0.$$

When $\ell=0$ and $\mu=0$, then $A=0$ and the imaginary part of the dominant quasinormal mode ($n=0$),
$$
\im{\omega}=\Order{eQ,\kappa^2},
$$
which cannot be computed within the simple approach presented in this section. However, Eq.~(\ref{kappaw}) allows us to conclude that whatever small mass $\mu>0$ stabilizes the scalar field, at least for sufficiently small values of its charge $e$.

For the Reissner-Nordström-de Sitter black hole the instability was observed only when $\ell=0$~and~$\mu=0$~\cite{Zhu:2014sya,Konoplya:2014lha}.
Therefore, when studying the parametric region of the scalar-field instability in the background of the dilaton-de Sitter black hole, we will represent here mostly the spherically symmetric ($\ell=0$) perturbations of the massless scalar field ($\mu=0$).

\section{Time-domain analysis}\label{sec:timedomain}

Here in order to integrate the wave equation~(\ref{radial_eq}) and analyze the spectrum of the perturbation, we shall use, first of all, the time-domain integration, which includes contribution from all modes.
The discretization scheme which we shall use was proposed in~\cite{Abdalla:2010nq}.
Defining $\Psi(r_\star,t)=\Psi(j\Delta r_\star, i\Delta t)=\Psi_{j,i}$, $V(r(r_\star))=V(j\Delta r_\star)=V_j$ and $\Phi(r(r_\star))=\Phi(j\Delta r_\star)=\Phi_j$, we can write down (\ref{radial_eq}) as
\begin{eqnarray}
&&-\frac{(\Psi_{j,i+1}-2\Psi_{j,i}+\Psi_{j,i-1})}{\Delta t^2}-2i\Phi_j\frac{(\Psi_{j,i+1}-\Psi_{j,i-1})}{2\Delta t} \nonumber \\ \nonumber
&&+\frac{(\Psi_{j+1,i}-2\Psi_{j,i}+\Psi_{j-1,i})}{\Delta r_\star^2}-V_j\Psi_{j,i}={\cal O}(\Delta t,\Delta r_\star).
\end{eqnarray}
The initial Gaussian wave-package has the form
$$\Psi(r_\star,t=0)=\exp\bigg[-{(r_\star-a)^2/2b^2}\bigg], \quad \Psi(r_\star,t<0)=0.$$
Then, the evolution of $\Psi$ can be described by the following expression
\begin{eqnarray}
\Psi_{j,i+1}=-{\frac{(1-i\Phi_j\Delta t)\Psi_{j,i-1}}{1+i\Phi_j\Delta t}}+{\frac{\Delta t^2}{\Delta r_\star^2}}
{\frac{\Psi_{j+1,i}+\Psi_{j-1,i}}{1+i\Phi_j\Delta t}} \nonumber\\\nonumber
+\bigg(2-2{\frac{\Delta t^2}{\Delta r_\star^2}}-\Delta t^2V_j\bigg){\frac{\Psi_{j,i}}{1+i\Phi_j\Delta t}}\;.
\end{eqnarray}
Following \cite{Abdalla:2010nq}, we choose the parameters $a=0$ and $b=\sqrt{10}$ in the Gaussian wave package and use
\begin{equation}
{\frac{\Delta t}{\Delta r_\star}}=\frac{1}{2}<1,
\end{equation}
making sure that $\Delta t$ is small enough to achieve the required precision of the profile.

In order to calculate $\re{\omega}$ we used the Prony method of fitting the time-domain profile data by superposition of damping exponents~\cite{Berti:2007dg}
\begin{equation}\label{damping-exponents}
\Psi(r,t)\simeq\sum_{i=1}^pC_ie^{-\imo\omega_i (t-t_0)}.
\end{equation}
We consider a late time period, which starts at $t_0$ and ends at $t=N\Delta t+t_0$, where $N$ is an integer and $N\geq2p-1$. Then the formula~(\ref{damping-exponents}) is valid for each value from the profile data:
\begin{equation}
x_n\equiv\Psi(r,n\Delta t+t_0)=\sum_{j=1}^pC_je^{-\imo\omega_j n\Delta t}=\sum_{j=1}^pC_jz_j^n.
\end{equation}
The Prony method allows us to find $z_i$ in terms of the known $x_n$ and, since $\Delta t$ is also known, to calculate the quasinormal frequencies $\omega_i$.

\begin{figure}
\resizebox{\linewidth}{!}{\includegraphics*{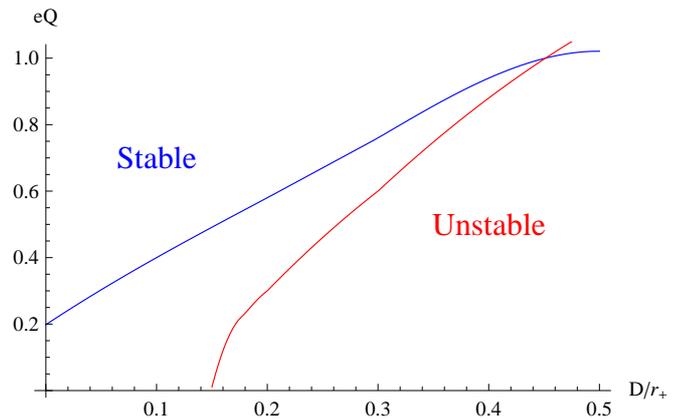}}
\caption{The region of (in)stability for $r_c=4r_+$ (blue, upper) and $r_c=3r_+$ (red, lower).}\label{fig:instability}
\end{figure}

\section{Bernstein spectral method}\label{sec:Bernstein}

First, we will discuss the Bernstein spectral method for asymptotically de Sitter spacetimes, when the purely outgoing wave is imposed at the de Sitter horizon.
Following~\cite{Fortuna:2020obg}, we introduce the compact coordinate, which is defined as follows:
\begin{equation}\label{compact}
u\equiv\frac{\frac{1}{r}-\frac{1}{r_c}}{\frac{1}{r_+}-\frac{1}{r_c}},
\end{equation}
and define the function $\psi(u)$, which is regular for $0\leq u\leq 1$ when $\omega$ is a quasinormal mode,
\begin{equation}\label{regularized}
\Psi(u)=u^{-\imo\Omega_c(\omega)}(1-u)^{-\imo\Omega_+(\omega)}\psi(u),
\end{equation}
where $\Omega_c(\omega)$ and $\Omega_+(\omega)$ are obtained from the characteristic equations at the singular points, $u=0$ and $u=1$, of the wavelike equation~(\ref{radial}).

In order to satisfy the quasinormal boundary conditions (\ref{QNBC}), we fix the values of $\Omega_c(\omega)$ and $\Omega_+(\omega)$, such that
\begin{equation}\label{BCfix}
\frac{d\Omega_c}{d\omega~}>0,\qquad\frac{d\Omega_+}{d\omega~}>0.
\end{equation}

We represent $\psi(u)$ as a sum
\begin{equation}\label{Bernsteinsum}
\psi(u)=\sum_{k=0}^NC_kB_k^N(u),
\end{equation}
where
$$B_k^N(u)\equiv\frac{N!}{k!(N-k)!}u^k(1-u)^{N-k}$$
are the Bernstein polynomials.

Substituting (\ref{regularized})~into~(\ref{radial}) and using a Chebyshev collocation grid of $N+1$ points,
$$u_p=\frac{1-\cos \frac{p\cdot\pi}{N}}{2}=\sin^2\frac{p\cdot\pi}{2N}, \qquad p=\overline{0,N},$$
we obtain a set of linear equations with respect to $C_k$, which has nontrivial solutions iff the corresponding coefficient matrix is singular. Since the elements of the coefficient matrix are polynomials (of degree 2) of $\omega$, the problem is reduced to the eigenvalue problem of a matrix pencil (of order 2) with respect to $\omega$, which can be solved numerically. Once the eigenvalue problem is solved, one can calculate the corresponding coefficients $C_k$ and explicitly determine the polynomial (\ref{Bernsteinsum}), which approximates the solution to the wave equation (\ref{radial}).

In order to exclude the spurious eigenvalues, which appear due to finiteness of the polynomial basis in~(\ref{Bernsteinsum}), we compare both the eigenfrequencies and corresponding approximating polynomials for different values of $N$. First, from each set of the solutions we take the eigenvalues that differ less than the required accuracy. Then, following~\cite{Konoplya:2022xid}, for each pair of the corresponding eigenfunction, $\psi^{(1)}$ and $\psi^{(2)}$, we calculate
$$1-\frac{|\langle \psi^{(1)}\;|\;\psi^{(2)} \rangle|^2}{||\psi^{(1)}||^2||\psi^{(2)}||^2}=\sin^2\alpha,$$
where $\alpha$ is the angle between the vectors $\psi^{(1)}$ and $\psi^{(2)}$ in the $L^2$-space. If all values of $\alpha$ are sufficiently small, we conclude that the obtained eigenvalues $\omega$ approximate the quasinormal frequencies, and better approximations correspond to larger $N$. The error estimation can be done by calculating the difference between the approximate eigenvalues of $\omega$, corresponding to different values of $N$.

Notice that, unlike \cite{Fortuna:2020obg}, we compare the polynomial approximations for the eigenfunctions without a normalization, so that the eigenfunctions, obtained for different $N$, have different complex constant prefactor.

It is a valuable feature of the Bernstein method that it can be similarly applied to asymptotically flat~($\Lambda=0$) and AdS black holes~($\Lambda<0$), which correspond to a qualitatively different boundary condition at infinity.
In this case the compact coordinate is introduced as follows
\begin{equation}\label{icompact}
u=\frac{r_+}{r}.
\end{equation}

Notice that although for $r_c\to\infty$ the quasinormal spectrum of the asymptotically de Sitter black hole approaches the one of the flat black hole and Eq.~(\ref{compact}) leads to the correct expression for the compact coordinate (\ref{icompact}). The prefactor in (\ref{regularized}) and the equation for the regular function $\psi(u)$ cannot be obtained by taking this limit. The reason is that the singular point $u=0$ becomes irregular for the asymptotically flat spacetime. The regular function $\psi(u)$ is defined as follows:
\begin{equation}\label{regularized-flat}
\Psi(u)=e^{\imo\Omega_c(\omega)/u}u^{\alpha(\omega)}(1-u)^{-\imo\Omega_+(\omega)}\psi(u),
\end{equation}
{where $\Omega_c(\omega)$, $\alpha(\omega)$, and $\Omega_+(\omega)$ are again determined by solving the characteristic equations, and the quasinormal boundary conditions imply that $\Omega_c(\omega)$ and $\Omega_+(\omega)$ satisfy Eq.~(\ref{BCfix}).\footnote{Similarly, the singular point $u=1$ becomes irregular for the extreme charge of the Reissner-Nordström-(A)(dS) black hole.}}

For the AdS black hole, the regular function $\psi(u)$ is introduced as
\begin{equation}\label{regularized-AdS}
\Psi(u)=u^2(1-u)^{-\imo\Omega_+(\omega)}\psi(u),
\end{equation}
so that the function $\Psi(u)$ obeys the Dirichlet boundary condition at spatial infinity ($u=0$).

After obtaining of the differential equation for the regular function $\psi(u)$, one applies the expansion~(\ref{Bernsteinsum}) for dS, flat, and AdS case, and finally applies the spectral method \cite{Jansen:2017oag,Fortuna:2020obg}.

We publicly share the Mathematica\textregistered{} package allowing one to obtain the equation for the regular function $\psi(u)$, defined by Eqs.~(\ref{regularized})~(de Sitter), (\ref{regularized-flat})~(flat), or (\ref{regularized-AdS})~(anti-de Sitter), solve the eigenvalue problem for the finite Bernstein polynomial basis, and compare different approximations~\cite{package}. Indeed, the region of instability shown in Fig.~\ref{fig:instability} is visually indistinguishable when plotted via the data from the time-domain integration or Bernstein spectral method.

\section{Numerical results}\label{sec:results}

\begin{table}
\begin{tabular}{|l|c|c|}
\hline
$eQ$ & Time-domain & Bernstein \\
\hline
$0.1 $ & $0.010792 + 0.000318 \imo$ & $0.01079212 + 0.00031822 \imo$ \\
$0.5 $ & $0.059873 + 0.003736 \imo$ & $0.05987310 + 0.00373586 \imo$ \\
$0.7 $ & $0.087104 + 0.001812 \imo$ & $0.08710412 + 0.00181189 \imo$ \\
$0.8 $ & $0.100198 - 0.000342 \imo$ & $0.10020300 - 0.00033985 \imo$ \\
$0.9 $ & $0.112675 - 0.002955 \imo$ & $0.11267481 - 0.00295525 \imo$ \\
$0.11$ & $0.135685 - 0.008578 \imo$ & $0.13572338 - 0.00857444 \imo$ \\
\hline
\end{tabular}
\caption{The fundamental quasinormal modes computed with the help of the time-domain integration and Bernstein spectral methods: $r_+=1$, $r_c=10$, $D=0.1$, $\ell=n=0$.}\label{tabl:1}
\end{table}

\begin{table}
\begin{tabular}{|l|c|c|}
\hline
$eQ$ & Time-domain & Bernstein \\
\hline
$0.003$ &$0.001773\! - 6.2\!\times\!\!10^{-7} \imo$& $0.00177273 - 7.7 \times 10^{-7} \imo$ \\
$0.01 $ & $0.005909 - 0.000008 \imo$&  $0.00590900 - 0.00000855 \imo$ \\
$0.05 $ & $0.029535 - 0.000214 \imo$ & $0.02953446-  0.00021305 \imo$ \\
$0.1  $ & $0.059004 - 0.000845 \imo$ & $0.05900394 - 0.00084449 \imo$ \\
$0.2  $ & $0.117516 - 0.003261 \imo$ & $0.11751631 - 0.00326156 \imo$ \\
$0.3  $ & $0.175176 - 0.006951 \imo$ & $0.17517755 - 0.00695443 \imo$\\
$0.4  $ & $0.231809 - 0.011563 \imo$ & $0.23181472 - 0.01156800 \imo$\\
$0.5  $ & $0.287384 - 0.016778 \imo$ & $0.28739890 - 0.01679517 \imo$ \\
\hline
\end{tabular}
\caption{The fundamental quasinormal modes computed with the help of the time-domain integration and Bernstein spectral methods: $r_+=1$, $r_c=2$, $D=0.1$, $\ell=n=0$.}\label{tabl:2}
\end{table}

\begin{table}
\begin{tabular}{|c|c|c|}
\hline
$eQ$ & Time-domain & Bernstein \\
\hline
$0.1$ & $0.295935 - 0.132693 \imo$ & $0.29594116 - 0.13269571 \imo$ \\
$0.1$ & $0.457250 - 0.137305 \imo$ & $0.45725754 - 0.13730920 \imo$ \\
$0.2$ & $0.217305 - 0.129911 \imo$ & $0.21729798 - 0.12990610 \imo$ \\
$0.2$ & $0.539761 - 0.139197 \imo$ & $0.53975494 - 0.13919893 \imo$ \\
$0.3$ & $0.140158 - 0.126766 \imo$ & $0.14015223 - 0.12675802 \imo$ \\
$0.3$ & $0.623383 - 0.140851 \imo$ & $0.62339015 - 0.14085650 \imo$ \\
$0.4$ & $0.064636 - 0.123247 \imo$ & $0.06463348 - 0.12323774 \imo$ \\
$0.4$ & $0.708055 - 0.142307 \imo$ & $0.70808605 - 0.14230319 \imo$ \\
$0.5$ & $0.009103 - 0.119386 \imo$ & $0.00910394 - 0.11937779 \imo$ \\
$0.5$ & $0.793728 - 0.143588 \imo$ & $0.79377577 - 0.14356120 \imo$ \\
\hline
\end{tabular}
\caption{The fundamental quasinormal modes computed with the help of the time-domain integration and Bernstein spectral methods: $r_+=1$, $r_c=2$, $D=0.1$, $\ell=1$, $n=0,1$.}\label{tabl:3}
\end{table}

We calculated quasinormal modes with the help of the two above methods: time-domain integration and the Bernstein spectral method. Both methods very well agree on the region of instability, though the time-domain integration is much more (at least two orders) time consuming. Indeed, the curves in Fig.~\ref{fig:instability} are visually indistinguishable when constructed on the data from either of the two methods. The larger is $r_{+}$ the larger is the stability region, that is, at the smaller $e Q$ the configuration is stabilized. The regime of tiny $e Q$ for the weakly charged Reissner-Nordström-de Sitter black hole is unstable unless $r_{+}$ is not larger than some critical value (see Fig.~6 in~\cite{Konoplya:2014lha}). The dilaton charge $D$ changes the situation: at a fixed $r_{+}$ and $r_c$ the system is stabilized at a larger $e Q$. In a similar fashion with the Reissner-Nordström-de Sitter solution~\cite{Konoplya:2014lha}) larger values of $r_{+}$ make the region of stability larger.

The data given in the tables shows that the fundamental quasinormal modes computed by the Bernstein spectral method are in excellent agrement with those extracted from the time-domain profiles. A tiny difference between them, must be interpreted in favor of the Bernstein spectral method, because in order to find quasinormal frequency from the time-domain profile, one needs not only to integrate the weave equation with enormous accuracy, but also accurately guess the beginning and end of the ringdown period, in order to avoid the mixture of the initial outburst and asymptotic tails.

From the Tables~\ref{tabl:1} and~\ref{tabl:2} one can see that the region of instability indeed can be determined with the help of the Bernstein spectral method, which allows us to calculate the dominant modes (either with positive or negative imaginary part) with good accuracy. In Table~\ref{tabl:3} we show that the method is also efficient for the accurate calculation of the overtones. This can be seen by comparing the obtained frequencies with those extracted from the time-domain profiles for larger multipole number, $\ell\geq1$, when the ringing occurs for sufficiently long time, allowing us to determine the dominant modes with higher accuracy.

With the help of the spectral method one can easily check that the instability occurs for the spherically symmetric perturbations ($\ell=0$) only. Using the Mathematica\textregistered{} package~\cite{package}, it is possible to find the accurate value of the scalar-field instability threshold (critical scalar-field charge) for any given black-hole parameters, $r_+$, $r_c$, and $D$.

\section{Nonoscillatory  modes of a neutral scalar field}\label{sec:dS}

Quasinormal modes of gravitational and other spin field perturbations of the Schwarzschild-de Sitter black holes
are known to consist from the two branches: the Schwarzschild-like modes deformed by the nonzero value of the cosmological constant and the purely imaginary (i.e., nonoscillatory) modes of the empty de Sitter spacetime \cite{Lopez-Ortega:2006aal} corrected by the nonzero mass of the black hole. When the black hole is small in comparison with the cosmological scale, the de Sitter modes become the least damping ones, thereby, dominating in a signal at late times and showing itself as exponential tails \cite{Konoplya:2022xid}. The de Sitter branch cannot be reproduced with the most frequently used WKB method \cite{Konoplya:2022gjp} and requires usage of accurate methods based on the converging procedures, such as the Frobenius expansion or Bernstein spectral method.

In \cite{Konoplya:2022kld} it was shown that the above result is general for small black holes in an arbitrary metric theory of gravity immersed in the de Sitter spacetime: the dominant quasinormal modes of the uncharged massless fields are purely imaginary, approaching the de Sitter modes according to the following universal law
\begin{equation}\label{dSmodes}
  \omega=-\imo\frac{\ell+\n+1-\delta_{\n0}\delta_{s0}}{r_c}\left(1-\frac{M}{r_c}\right)+\Order{\frac{1}{r_c^3}},
\end{equation}
where $s=0$ for the scalar field and $\n=0,1,2\ldots$ enumerates the purely imaginary modes.

\begin{table}
\begin{tabular}{|c|l|l|l|}
\hline
$eQ=0$             & $D=0.1$        & $D=0.2$        & analytic      \\
\hline
$\ell=1$, $r_c=10$ & $-0.09585\imo$ & $-0.09684\imo$ & $-0.0950\imo$ \\
$\ell=1$, $r_c=20$ & $-0.04898\imo$ & $-0.04923\imo$ & $-0.0488\imo$ \\
\hline
$\ell=2$, $r_c=10$ & $-0.1918\imo$  & $-0.1938\imo$  & $-0.1901\imo$ \\
$\ell=2$, $r_c=20$ & $-0.098\imo$   & $-0.099\imo$   & $-0.0975\imo$ \\
\hline
\end{tabular}
\caption{Comparison of the purely imaginary (dominant, $\n=0$) mode for the scalar field, obtained using the Bernstein spectral method, and the analytic formula (\ref{dSmodes}) in units of the horizon radius ($r_+=1$).}\label{tabl:comp}
\end{table}

In Table~\ref{tabl:comp} we compare the above analytic formula (\ref{dSmodes}) and the numerical values obtained with the help of the Bernstein spectral method. We can see that the Bernstein method is very efficient not only for finding purely imaginary algebraically special modes, but also for purely imaginary modes satisfying the quasinormal boundary condition.

\section{Conclusions}\label{sec:conclusions}

Here we represented the improved user-friendly Mathematica\textregistered{} procedure for finding quasinormal modes of a wide class of black holes with the help of the Bernstein spectral method.
The code we are sharing is automatic at a great extent and can be used for a wide class of wave equations allowing for asymptotically flat, de Sitter, or anti-de Sitter boundary conditions. By applying the method to the charged scalar field perturbations around a charged dilatonic asymptotically de Sitter black hole, we show that the method is very efficient when studying the regime of instability. We have checked our results with the help of the (less economic) time-domain integration method and achieved excellent agreement between both methods. We have shown that the instability for a dilaton-de Sitter black hole has superradiant nature, in a similar way with the Reissner-Nordström-de Sitter case.

The difficulty when using the Bernstein spectral method is related to the great number of roots of the matrices which are the bigger, the more overtones one wish to find, or the higher accuracy of the particular modes are required. Then, the question is which roots represent quasinormal frequencies and which are the numerical artifacts. Our code includes the comparison for matrices of various sizes, allowing to discard the numerical artifacts. The code is automatic and does not require essential modifications when applying it to this or that black-hole perturbation equations.

The Bernstein spectral method could be a relatively quick way to detect instability of black holes or find purely imaginary modes. In principle, it could be extended to a system of chained wave equations and at least to some axially symmetric black holes. We believe that future publications will study these questions.

\begin{acknowledgments}
The authors acknowledge support of the grant 19-03950S by Czech Science Foundation (GAČR).
A. Z. was supported by Conselho Nacional de Desenvolvimento Científico e Tecnológico (CNPq).
\end{acknowledgments}

\end{document}